\input amstex
\documentstyle{amsppt}
\topmatter
\title An L-A pair for the Apel'rot system and a new integrable case for the
Euler-Poisson equations on $so(4)\times so(4)$
\endtitle
\rightheadtext {An L-A pair for the Apelrot system}
\leftheadtext {Vladimir Dragovi\'c, Borislav Gaji\'c}
\endtopmatter
\document
\def\Ree{\operatorname {Re}}
\def\Det{\operatorname {Det}}
\def\Tr{\operatorname {Tr}}
\def\Im{\operatorname {Im}}
\def\Jac{\operatorname {Jac}}
\TagsOnRight

\baselineskip=20pt

{\bf Vladimir Dragovi\' c} and {\bf Borislav Gaji\' c}

{Mathematical Institute SANU}, {Kneza Mihaila 35, 11000 Beograd}, {Yugoslavia}

{E-mails:} {vladad{\@}mi.sanu.ac.yu} and {gajab{\@}mi.sanu.ac.yu}
\vskip 1cm
{\eightpoint
We present an L-A pair for the Apel'rot case of a heavy rigid 
3-di\-men\-si\-o\-nal body. Using it we give an algebro-geometric integration
procedure. Generalizing this L-A pair, we obtain a new completely integrable 
case of the Euler-Poisson equations in dimension four. Explicit formulae for
integrals which are in involution are given. This system is a counterexample to
one well known Ratiu's theorem. Corrected version of this classification
theorem is proved.}

\

{\bf 1. Introduction}

The rotations of a heavy rigid body fixed at a point are described by the
Euler-Poisson equations. It is well known that these equations are integrable
in Euler, Lagrange and Kovalevskaya cases. For a fixed value of one of
the integrals, there are additional integrable cases, for example 
Goryachev-Chaplygin and Apel'rot (see [{\bf 1}], [{\bf 5}]). L-A pairs were known 
for all these problems except the last one (see [{\bf 4}], [{\bf 7}], 
[{\bf 10}], [{\bf 11]}). 
In this paper, in section 2, 
we present an L-A pair for the Apel'rot case. (Almost the same L-A pair
serves in the Apel'rot gyrostat case.) Using it, we give an 
algebro-geometric integration procedure in section 3. The spectral curve is 
reducible and it consists of a sphere and a torus. The evolution of the pole
divisor of the eigen-function of the $L$ matrix is linearized on the Jacobian
of the torus. It leeds to the solution of the initial problem up to rotation.
For a complete solution one needs one step more: to integrate the Riccati
equation with double-periodical coefficients. Generalizing this L-A pair, 
we obtain, in section 4, a new completely integrable system of a 
four-dimensional rigid 
body motion. This system is integrable even without reduction to any invariant 
submanifold. Explicit formulae for integrals wich are in involution are
given.

It turns out that the L-A pair, which we use, is of the form analysed by Ratiu 
and van Moerbeke in the study of the generalized Lagrange case and of the 
completely symmetrical case in [{\bf 9}], [{\bf 10}]. Our system is a 
counterexample to Ratiu's well known theorem (see [{\bf 9}], [{\bf 13}]). 
This theorem claims that the Lagrange and the completely symmetric case have 
exhausted the list of the Euler-Poisson equations which are equivalent to 
an L-A pair of that form. We conclude the paper by giving a corrected version 
of this classification theorem.

\

{\bf 2. Apel'rot case}

The equations of rotation of a heavy rigid body fixed at a point in a
moving frame are:
$$
\frac{d}{dt}{\vec M}={\vec M}\times\vec\omega+\vec\gamma\times \vec r_C,\qquad 
\frac{d}{dt}{\vec\gamma}=\vec\gamma\times\vec\omega,
\tag1
$$
where $\vec\omega$ is the vector of the angular velocity, $\vec M=I\vec\omega$ 
is the kinetic momentum, $I$ is the inertia operator, $\vec\gamma$ is the 
unit vertical vector, and $\vec r_C=(x_0, y_0, z_0)$ is the radius vector of 
the mass center according to the fixed point. One can choose
the moving frame such that the inertia operator is diagonal, $I=diag(I_1, I_2,
I_3)$. The equations (1) have the following three integrals of motion 
(see [{\bf 3}]):
$$
\aligned
F_1&=\frac 12\langle I\vec\omega, \vec\omega\rangle+\langle \vec\gamma,
\vec r_C\rangle,\\
F_2&=\langle\vec\gamma, \vec\gamma\rangle(=1),\\
F_3&=\langle I\vec\omega, \vec\gamma\rangle.
\endaligned
$$
For complete integrability, we need one integral more. In 1894 Apel'rot
noticed (see [{\bf 1}], [{\bf 5}]) that under the additional conditions:
$$
\aligned
&i)\qquad y_0=0\\
&ii)\qquad x_0\sqrt{I_1(I_2-I_3)}+z_0\sqrt{I_3(I_1-I_2)}=0
\endaligned
\tag2
$$
the equations (1) became integrable on the hypersurface given with:
$$
I_1x_0\omega_1+I_3z_0\omega_3=0.
\tag3
$$
A geometric interpretation of these conditions was given by Zhukovski [{\bf
6}],[{\bf 14}].

Using standard isomorphism between Lie algebras $R^3$ and $so(3)$ 
(notation: $\vec A\in R^3\mapsto A\in so(3)$), (see [{\bf 2}], [{\bf 10}]), 
the equations
(1) can be transformed to the following equations on $so(3)\times so(3)$:
$$
\frac{dM}{dt}=[M,\omega]+[\gamma,r_c],\qquad
\frac{d\gamma}{dt}=[\gamma,\omega]
\tag4
$$
\proclaim{Theorem 1}
Under the Apel'rot conditions (2), the equations (4) are equivalent to:
$$
\frac d{dt}(\lambda^2C+\lambda M+\gamma)=
[\lambda^2C+\lambda M+\gamma,\omega+\lambda
r_C],
\tag5
$$
where $\lambda$ is a spectral parameter and $C\in so(3)$ is a constant 
matrix given by
$$
C=I_2 r_C.
\tag6
$$
\endproclaim
\demo{Proof}
The equations (4) are equivalent to (5) if and only if 
$$
[C,r_C]=0 ,\qquad [C,\omega]+[M, r_C]=0.
$$
The first equation is satisfied by (6). The second one can be reduced to the
equation:
$$
(I_1-I_2)\omega_1z_0+(I_2-I_3)\omega_3x_0=0.
\tag7
$$
But (7) is a consequence of the Apel'rot conditions (2) and the equation of
the hypersurface (3).
\enddemo

\

{\bf Note.} A gyrostat is a system which consists of two rigid bodies. The
second one, the gyroscope, is axialy-symmetric and it rotates with a constant
angular velocity about its symmetry axis which is fixed in the first rigid
body. The equations of motion of that system are of the form (4) with 
$$
M=I\omega+P
$$
where $P$ is a constant skew-symmetric matrix, which represents the kinetic
momentum of the gyroscope. 
Sretenskiy in [{\bf 12}] proved that, if Apel'rot conditions (2) are satisfied, 
and $P_2=0$, then the equations of motion are integrable on the hypersurface 
given by:
$$
(I_1-I_2)\omega_1 z_0+(I_2-I_3)\omega_3 x_0-(P_3x_0-P_1z_0)=0.
$$
It can easely be proved that: 
$$
L(\lambda)=\lambda^2C+\lambda M+\gamma;\qquad A(\lambda)=\omega+\lambda r_C,
$$
form the L-A pair for the Apel'rot gyrostat.

\

An L-A pair of the form
$$
L(\lambda)=\lambda^2C+\lambda M+\gamma,\qquad A(\lambda)=\omega+\lambda r_C
\tag8
$$
was used in [{\bf 10}] for the Lagrange case (of course, the matrices $r_C$ and
$C$ from [{\bf 10}] are different from the matrices $r_C$ and 
$C$ in this paper. Also, the conditions on the inertia operator are 
not the same).

\

{\bf 3. Integration of the Apel'rot case}

The next steps in algebro-geometric integration of the 
Apel'rot case follow the procedure from [{\bf 10}].

As usual, in a direct problem one corresponds to $L(\lambda)$ some
algebro-geometric data with simple evolution in time. After the integration
of this evolution, in the inverse part, one has to reconstruct the matrix
elements of $L(\lambda)$, starting from the algebro-geometric data as known
functions of time. The final part is the integration of the equations of
motion of the initial problem by use of the known matrix entries of 
$L(\lambda)$.
A specific characteristic of the problem presented here, is that the algebro-geometric
procedure gives a solution of the equations of motion of the Apel'rot case
only up to rotations. For the full solution of the problem, we need one
integration more to determine the angle variable $\phi_x=\arg x$ - see 
theorem 5 bellow.

Let:
$$
\alpha=\frac{x_0}{\sqrt{x_0^2+z_0^2}}\qquad
\beta=\frac{z_0}{\sqrt{x_0^2+z_0^2}}.
\tag9
$$
The matrix
$$
U=\left[
\matrix
\alpha& \frac{i\beta}{\sqrt2}&\frac{\beta}{\sqrt2}\\
0&\frac{1}{\sqrt2}&\frac{i}{\sqrt2}\\
\beta&\frac{-i\alpha}{\sqrt2}&\frac{-\alpha}{\sqrt2}
\endmatrix
\right]
$$
diagonalises matrix $C$. In the new basis, the matrix 
$L(\lambda)=C\lambda^2+M\lambda+\gamma$
transformes into:
$$\tilde{L}=U^{-1}LU=
\left[
\matrix
0&\Delta&i\Delta^{*}\\
-\Delta^{*}&-\Omega&0\\
i\Delta&0&\Omega
\endmatrix
\right]
$$
where:
$$
\aligned
\Delta&=y+\lambda x;\quad \Delta^*=\bar{y}+\lambda\bar{x};\\
y&=\frac1{\sqrt{2}}(\beta\gamma_1-\alpha\gamma_3-i\gamma_2);\quad  
x=\frac1{\sqrt{2}}(\beta M_1-\alpha M_3-iM_2)\\
\Omega&=-i\left[\alpha(C_1\lambda^2+M_1\lambda+\gamma_1)+
\beta(C_3\lambda^2+M_3\lambda+\gamma_3)\right]\\
&=-i\left[\alpha(C_1\lambda^2+\gamma_1)+
\beta(C_3\lambda^2+\gamma_3)\right]
\endaligned
\tag{10}
$$
and $C_1=I_2x_0;\ C_3=I_2z_0$, using (3).

The spectral curve is defined by:
$$
\Gamma: p(\mu,\lambda):=\det(L(\lambda)-\mu E)=0,
$$
where
$$
p(\mu,\lambda)=-\mu(\mu^2-\Omega^2+2\Delta\Delta^*).
\tag{11}
$$

The spectral curve $\Gamma$ is reducible and it consists of two components: 
the sphere $\Gamma_1$ given by $\mu=0$, and the torus $\Gamma_2$ :
$$
\mu^2=\Omega^2-2\Delta\Delta^*=:P_4(\lambda).
\tag{12}
$$

The coefficients of the spectral polynomial (12) are integrals of the motion. 
Substituting expressions (10) in (11), the following form of the equation 
of the spectral curve is obtained:
$$
p(\mu, \lambda)=-\mu(\mu^2+A\lambda^4+B\lambda^3+D\lambda^2+E\lambda+F),
\tag {13}
$$
where
$$
\aligned
A&=I_2^2(x_0^2+z_0^2)\\
B&=2I_2(M_1x_0+M_3z_0)(=0)\\
D&=2I_2\left(\frac{M_1^2}{2I_2}+\frac{M_2^2}{2I_2}+\frac{M_3^2}{2I_2}+
x_0\gamma_1+z_0\gamma_3\right)\\
E&=2(M_1\gamma_1+M_2\gamma_2+M_3\gamma_3)\\
F&=\gamma_1^2+\gamma_2^2+\gamma_3^2(=1)
\endaligned
\tag{14}
$$
Using the Apel'rot conditions, the expression for $D$ can be simplified:
$$
D=2I_2\left(\frac{M_1^2}{2I_1}+\frac{M_2^2}{2I_2}+\frac{M_3^2}{2I_3}+
x_0\gamma_1+z_0\gamma_3\right). 
$$
This is the energy integral. So, in this case the L-A pair (5) gives 
four integrals of the motion, which is enough for 
integrability.

Let $(f_1, f_2, f_3)^T$ denote an eigenvector of the matrix $L(\lambda)$, 
which corresponds to the eigenvalue $\mu$. Fix a normalizing condition 
$f_1=1$. Then:
$$
f_2=-\frac{\Delta^*}{\Omega+\mu},\qquad f_3=-\frac{i\Delta}{\Omega-\mu}.
\tag{15}
$$
The restrictions to $\Gamma_2$ are:
$$
f_2=-\frac{\Omega-\mu}{2\Delta},\qquad f_3=\frac{\Omega+\mu}{2i\Delta^*}.
$$ 
The relation between the divisors on $\Gamma_2$:
$$
(f_2)+(f_3)=0.
$$
is a consequence of $f_2\cdot f_3=\frac i2$.

From (12) the divisors of $\lambda$ and $\mu$ on $\Gamma_2$ are:
$$
\aligned
(\lambda)&=-P_1-P_2+R_1+R_2\\
(\mu)&=-2P_1-2P_2+(\mu)_0,\quad \deg(\mu)_0=4
\endaligned
$$
where $P_1$ and $P_2$ are the points on $\Gamma_2$ over $\lambda=\infty$.
From (12), we obtain the developments in neighborhoods of $P_1$ and $P_2$:
$$
\frac{\mu}{\lambda^2}=\cases
iI_2\sqrt{x_0^2+z_0^2}+O(\lambda^{-2}),\quad\text{\rm in a neighborhood of}
\ P_1\\
-iI_2\sqrt{x_0^2+z_0^2}+O(\lambda^{-2}),\quad\text{\rm in a neighborhood of}
\ P_2.
\endcases
$$
From (15), it follows that $f_2$ has a simple pole in $P_1$ and a simple zero 
in $P_2$; $f_3$ has a simple zero in $P_1$ and a simple pole in $P_2$.
In the affine part, $f_2$ has a simple pole in $\nu$ given by 
$$
\nu:\ \Delta=0,\ \Omega+\mu=0,\quad \text{\rm i.e.}\quad \nu_\lambda=
-\frac yx,\quad \nu_\mu=-\Omega\mid_{\lambda=-\frac yx}.
$$
It has a simple zero in $\bar{\nu}=(\bar{\nu}_\lambda, \bar{\nu}_\mu)$, where
$\bar{\nu}_\lambda$ denotes the complex conjugate of $\nu_\lambda$.

The previous consideration gives us:
\proclaim{Lemma 1} The divisors of $f_2$ and $f_3$ on $\Gamma_2$ are:
$$
\aligned
(f_2)&=-P_1+P_2-\nu+\bar{\nu}\\
(f_3)&=P_1-P_2+\nu-\bar{\nu}
\endaligned
$$
\endproclaim
Now we are going to analyse the converse problem. Suppose the evolution 
in time of the point $\nu$ is known (see theorem 4 bellow). For reconstructing 
the matrix $L(\lambda)$, one needs $x=|x|e^{i\arg x},\ y=|y|e^{i\arg y}$.
\proclaim {Theorem 2}
The point $\nu\in\Gamma_2$ and the initial conditions for $M$ and $\gamma$ 
determine $|x|,\ |y|$ and $\arg y-\arg x$, where $x$ and $y$ are given by (10).
\endproclaim
\demo{Proof}
Let $\nu$ have coordinates $\nu_\mu$ and $\nu_\lambda$. From the expressions 
for $\nu$
$$
\nu_\lambda=-\frac yx;\ \nu_\mu=-\Omega\mid_{\lambda=-\frac yx}
$$
and expression (10) for $\Omega$, it follows:
$$
\alpha\gamma_1+\beta\gamma_3=-i\nu_\mu-I_2\nu_\lambda^2\sqrt{x_0^2+z_0^2}.
\tag{16}
$$
From (10) and (16) we can determine $\Omega$ as a polynomial 
of $\lambda$:
$$
\Omega=-i[I_2\lambda^2\sqrt{x_0^2+z_0^2}-i\nu_\mu-I_2\nu_\lambda^2
\sqrt{x_0^2+z_0^2}].
\tag{17}
$$
In the expression for the spectral curve $\mu^2=\Omega^2-2\Delta\Delta^*$ 
understood as a polynomial in $\lambda$, the coefficients are integrals of the 
motion (they are determined and given by (14)).

So
$$
\mu^2-\Omega^2=-2\Delta\Delta^*=-2(|y|^2+(y\bar{x}+x\bar{y})\lambda+|x|^2
\lambda^2)
$$
is a known polynomial and we can determine $|x|^2, |y|^2$ and $\arg y-\arg x=
\arg\left(\frac{y}{x}\right)$. 
\enddemo

The previous theorem gives us a possibillity for integration of the Apel'rot 
case if we know the motion of the point $\nu$. As we will show, the motion of 
the point $\nu$ is linearized on the Jacobian $\Jac(\Gamma_2)\cong\Gamma_2$.

In all propositions untill the end of this section, we will assume that the 
Apel'rot conditions are satisfied.
\proclaim{Theorem 3} If $\nu_\lambda$ and $\nu_\mu$ are coordinates of point 
$\nu\in\Gamma_2$, then:
$$
\frac{d}{dt}\nu_\lambda=\frac 1{I_2}\nu_\mu
$$
\endproclaim
\proclaim{Lemma 2} The functions $x$ and $y$ satisfy the system 
$$
\aligned
\frac {dx}{dt}&=\left(\frac1{I_1}-\frac1{I_2}\right) \frac{iM_1}
{\alpha \sqrt{2}}x+i\sqrt{x_0^2+z_0^2}y\\
\frac{dy}{dt}&=-\frac{i(\alpha\gamma_1+\beta\gamma_3)}{I_2}x+
\frac {iM_1} {\alpha\sqrt 2}\left(\frac 1{I_1}-\frac 1{I_2}\right)y
\endaligned
\tag{18}
$$
\endproclaim
\demo{Proof} From the Apelrot conditions (2) we can get
$$
\frac{\beta^2}{\alpha^2}\left(\frac1{I_1}-\frac 1{I_2}\right)=\frac 1{I_2}-
\frac 1{I_3}.
\tag{19}
$$
From (19), (3), and the identity $\alpha^2+\beta^2=1$ one obtaines
$$
\aligned
\frac{dx}{dt}&=\left(\frac1{I_1}-\frac1{I_3}\right) \frac{M_1}
{\alpha \sqrt{2}}(\alpha M_2-iM_3)+i\sqrt{x_0^2+z_0^2}y\\
\frac{dy}{dt}&=-\frac{i(\alpha\gamma_1+\beta\gamma_3)}{I_2}x+
\frac {iM_1} {\alpha\sqrt 2}\left(\frac 1{I_1}-\frac 1{I_2}\right)y.
\endaligned
$$
The claim follows applying (3).
\enddemo
As a consequence we get
\proclaim{Lemma 3} The function $\nu_\lambda=-\frac yx$ satisfies the equation:
$$
\frac{d}{dt}\left(-\frac yx\right)=\frac1{x^2}\left(\frac {dx}{dt}y-\frac{dy}
{dt}x\right)=
\frac1{x^2}\left[i\sqrt{x_0^2+z_0^2}y^2+
\frac{i(\alpha\gamma_1+\beta\gamma_3)}{I_2}x^2\right].
$$
\endproclaim
\demo{Proof of theorem 3} 
The divisor $\nu$ is given by $\nu_\lambda=-\frac yx,\ 
\nu_\mu=-\Omega\mid_{\lambda=-\frac yx}$. From lemma 2 and lemma 3,
we have:
$$
\aligned
\frac{d}{dt}\nu_\lambda=&\frac{d}{dt}\left(-\frac yx\right)=
\frac1{x^2}\left(\frac {dx}{dt}y-x\frac {dy}{dt}\right)=
\frac 1{x^2}\left[i\sqrt{x_0^2+z_0^2}y^2+\frac{i(\alpha\gamma_1+\beta\gamma_3)}
{I_2}x^2\right]=\\
&\frac 1{I_2}\left(-\Omega\mid_{\lambda=-\frac yx}\right)=\frac 1{I_2}\nu_\mu.
\endaligned
\tag{20}
$$
\enddemo

Since $\Gamma_2$ is a curve of genus 1, there is only one linearly independent 
holomorphic differential 
$$
\frac{d\lambda}{\mu}=\frac{d\lambda}
{\sqrt{\Omega^2-2\Delta\Delta^*}},
$$
on it.

So any linear flow on $\Jac(\Gamma_2)$ is of the form 
$\frac{d\lambda}{dt}=const\cdot\mu$.

A consequence of (20) is that the flow of the point $\nu$ is linear on 
$\Jac(\Gamma_2)$. 
We have the next theorem:
\proclaim {Theorem 4}
The integration of the motion of the point $\nu$ reduces to the inversion 
of the elliptical integral
$$
\int_{\nu_0}^\nu\frac{d\lambda}{\sqrt{\Omega^2-2\Delta\Delta^*}}=
\frac 1{I_2}t.
$$
\endproclaim

From theorem 2 and expression (10) for $x$ and $y$, it follows that, if we 
know the motion of the point $\nu$, in order to determine the matrix
$L(\lambda)$ as a function of 
time, we need $\arg x$ as a function of time. Let us denote by $\phi_x=\arg
x$, and $u=tg \frac{\phi_x}{2}$.
\proclaim{Theorem 5}
The function $u(t)$ satisfies the Riccati equation:
$$
\frac{du}{dt}=[(f(t)-g(t)]u^2+[f(t)+g(t)],
\tag{21}
$$
where
$$
\aligned
f(t)&=\frac K{2|x|^2}\quad g(t)=\frac{Q|x|}{2}\\
K&=\frac{\langle M, \gamma\rangle}{2\sqrt{x_0^2+z_0^2}}\\ 
Q&=\frac {\beta}{\alpha}\sqrt{2}\left(\frac 1I_2-\frac 1I_1\right)
\endaligned
\tag{22}
$$
and $|x|$ is a known function of time.
\endproclaim

\demo{Proof} 
From
$$
tg \phi_x=\frac{\Im x}{\Ree x}=\frac{-M_2}{\beta M_1-\alpha M_3},
$$
by differentiating and using the equations of motion (4), we get:
$$
\frac{d\phi_x}{dt}\frac1{\cos^2 \phi_x}=\frac{-M_1\langle M, M \rangle
\frac 1\alpha
\left(\frac 1I_2-\frac 1I_1\right)+\frac{\langle M, \gamma\rangle}
{\sqrt{x_0^2+z_0^2}}}{\langle M, M\rangle-M_2^2}.
\tag{23}
$$
From the system
$$
\aligned
&\alpha M_1+\beta M_3=0\\
&\beta M_1-\alpha M_3=\sqrt{2}\Ree x=\sqrt{2}|x|\cos \phi_x,
\endaligned
$$
$M_1$ can be expressed as a function of $|x|$, and $\phi_x$
$$
M_1=\sqrt{2}\beta |x|\cos \phi_x.
$$
Also, from (10) we have:
$$
M_2=-\sqrt{2}\Im x=-\sqrt{2}|x|\sin \phi_x.
$$
From (10), using (3) we have that $2|x|^2=\langle M, M\rangle$. 
Substituting the last three expressions into (23) gives
$$
\frac{d\phi_x}{dt}=\frac{K-Q|x|^3\cos\phi_x}{|x|^2}.
\tag{24}
$$
By the change of variables $u=tg \frac{\phi_x}{2}$ the equation (24) takes 
the form (21).
\enddemo

The classical integration procedure in the so-called Hesse coordinates [{\bf
5}] also yields the Riccati equation (see [{\bf 5}]). By Nekrasov 
(see [{\bf 5}], [{\bf 8}]) it was 
reduced to a second order linear differential equation with double-periodical 
coefficients.

\

{\bf 4. A new integrable case on $so(4)\times so(4)$}

The equations of rotations of a heavy $n$-dimensional rigid body fixed at
a point on $so(n)\times so(n)$ are given in [{\bf 9}]. In the moving frame, 
these equations are:
$$
\frac{dM}{dt}=[M,\Omega]+[\Gamma, X],\qquad \frac{d\Gamma}{dt}=
[\Gamma, \Omega].
\tag{25}
$$
$M=I\Omega+\Omega I\in so(n)$ is the kinetic momentum, $\Omega\in so(n)$ is the
angular velocity, $I$ is a symmetric $n\times n$ matrix, $\Gamma
\in so(n)$, and $X\in so(n)$ is a given constant matrix. By choosing the
moving frame with $I$ diagonal, $I=diag(I_1,...,I_n)$, we have
$M_{ij}=(I_i+I_j)\Omega_{ij}$, and $I_i+I_j$ are the principal inertia
momenta. 

In [{\bf 9}], the Lagrange case was defined by $I_1=I_2=a;\ I_3=\dots=I_n=b,\ 
X_{12}\ne 0,\ X_{ij}=0,\ i,j\ne 1,2,\ i<j$. The completely symmetric case 
was defined there by
$I_1=\dots=I_n=a$, where $X\in so(n)$ is an arbitrary constant matrix.
It was shown in [{\bf 9}] that the equations (25) in these cases could be 
represented by the following L-A pair:
$$
\frac d{dt} (\lambda^2C+\lambda M+\Gamma)=[\lambda^2C+\lambda M+\Gamma, \lambda
X+\Omega],
\tag{26}
$$
where in the Lagrange case $C=(a+b)X$, and in the symmetric case $C=aX$.

The theorem 4.1 in [{\bf 9}], states that there are no other cases of
n-dimensional rigid bodies, with the equations equivalent to (26).

However, let us consider a 4-dimensional rigid body with the conditions:
$$
\aligned
I_1=I_2=a,\ &I_3=I_4=b, \\
X_{12}\ne 0,\ X_{34}\ne0,\ X_{ij}=0,\ &i,j\ne \{(1,2),(3,4)\},\ i<j.
\endaligned
\tag{27}
$$
Note that this is neither the Lagrange nor the completely symmetrical case. 
The next theorem shows that this example satisfies all assumptions of the
theorem 4.1 from [{\bf 9}]. So, the system (27) is a counterexample to that 
theorem.
\proclaim {Theorem 6}
The equations (25) for a 4-dimensional rigid body which satisfies (27) are
equivalent to (26) with the matrix $C=(a+b)X$. 
\endproclaim
\demo{Proof}
We have $[C, X]=0$ from (27). Also, $[C, \Omega]+[M, X]=0$, so (26) is 
equivalent to (25).
\enddemo

One can naturaly consider (25) for $n=4$ as an equation on the semidirect
product $so(4)\times so(4)$ with the following Poisson structure on the
orbits of the coadjoint action (see [{\bf 9}]):
$$
\aligned
\{\tilde f,\tilde g\}(\mu, \nu)&=-\mu([d_1f(\mu, \nu),d_1g(\mu, \nu)])\\
&-\nu([d_1f(\mu, \nu),d_2g(\mu, \nu)])\\
&-\nu([d_2f(\mu, \nu),d_1g(\mu, \nu)]),
\endaligned
$$
where ${\tilde f}, {\tilde g}$ are restrictions of the functions $f$ and $g$ 
to the orbit of the coadjoint action, and $d_if$ are the partial 
derivatives of $df$. From the L-A pair representation (26), we have that 
integrals of the motion for the system given by (27), are the coefficients in
the $\lambda$ polynomials $\Tr(\lambda^2C+\lambda M+\Gamma)^2$, 
and $(\Det(\lambda^2C+\lambda M+\Gamma))^{1/2}$.
Using the same arguments as in [{\bf 9}], it can be proved that
the system given by (27) has the following four Casimir functions:
$$
\aligned
J_1&=M_{34}\Gamma_{12}+M_{12}\Gamma_{34}+M_{14}\Gamma_{23}+M_{23}\Gamma_{14}-
M_{24}\Gamma_{13}-M_{13}\Gamma_{24}\\
J_2&=\Gamma_{34}\Gamma_{12}+\Gamma_{23}\Gamma_{14}-\Gamma_{13}\Gamma_{24}\\
J_3&=\Gamma_{12}^2+\Gamma_{13}^2+\Gamma_{14}^2+\Gamma_{23}^2+\Gamma_{24}^2+
\Gamma_{34}^2\\
J_4&=M_{12}\Gamma_{12}+M_{13}\Gamma_{13}+M_{14}\Gamma_{14}+M_{23}\Gamma_{23}+
M_{24}\Gamma_{24}+M_{34}\Gamma_{34}.
\endaligned
$$
Also, it follows that this system has 
four integrals:
$$
\aligned
F_1&=C_{12}M_{12}+C_{34}M_{34}\\
F_2&=C_{34}M_{12}+C_{12}M_{34}\\
F_3&=M_{12}M_{34}+M_{23}M_{14}-M_{13}M_{24}+C_{34}\Gamma_{12}+C_{12}\Gamma_{34}\\
F_4&=M_{12}^2+M_{13}^2+M_{14}^2+M_{23}^2+M_{24}^2+M_{34}^2+
2C_{12}\Gamma_{12}+2C_{34}\Gamma_{34},
\endaligned
$$
which are in involution.
Thus, we have
\proclaim {Proposition} The system (27) is completely integrable.
\endproclaim

Except the three cases mentioned above, there are no other systems of the 
type (25) which are equivalent to (26). This is proved in the following:
\proclaim{Theorem 7} 
Let us suppose that $X_{12}\ne 0$. The Euler-Poisson equations (25) can be
written in the form (26) if and only if the equations (25) describe:

a) for $n=4$, the motion of the Lagrange top, the completely symmetric top or
a rigid body which satisfies (27).

b) for $n\ne4$, the motion of the Lagrange or completely symmetric top.
\endproclaim
\demo{Proof}
The equations (26) are equivalent to (25), if and only if $[C,X]=0$ and 
$[C,\Omega]+[M,X]=0$. The second relation is equivalent to 
$$
C_{ij}=(I_i+I_k)X_{ij},\qquad
C_{ij}=(I_j+I_k)X_{ij},\ \ k\ne i,j.
\tag{28}
$$
From $C_{12}\ne0$ and (28), for $i=1, j=2$, it follows that $I_1=I_2$, 
$I_k=I_l$ for all $k,l\ne 1,2$. Let us fix $i, j$ different from 1 and 2. 
From (28), we have $I_i=I_j, I_k=I_l$, for all $k,l\ne i,j$. For $n\ne 4$, 
this means that $I_1=\dots=I_n$ and $X$ is an arbitrary matrix, or 
$I_1=I_2=a$, $I_3=\dots=I_n$ and $X_{ij}=0$ for $i,j\ne 1,2$. For $n=4$, we 
get one more case which satisfies
the conditions (27). For such $C_{ij}$, the relation $[C, X]=0$ is satisfied.
\enddemo

\

{\bf Acknowledgment.} This research is supported by Ministry of Science and
Technology of Serbia, Project 04M03. We are gratefull to B. Jovanovi\'{c} for
stimulating discussions.

\

{\bf References}

\item{1} Apel'rot G.G.: The problem of motion of a rigid body 
about a fixed point. {\it Uchenye Zap. Mosk. Univ. Otdel. Fiz. Mat. Nauk} 
No. 11, (1894), 1-112.

\item{2} Arnol'd V. I.: {\it Mathematical methods of classical mechanics},
(Moscow: Nauka, 1989 [in Russian, 3-rd edition]). 

\item{3} Arnol'd V. I., Kozlov V. V., Neishtadt A. I.: {\it Mathematical 
aspects of classical and celestial mechanics/ in Dynamical systems III}, 
(Berlin: Springer-Verlag, 1988).

\item{4} Belokolos E.D., Bobenko A.I., Enol'skii V.Z., Its A.R., 
Matveev V.B.: {\it Algebro-geometric approuch to nonlinear integrable 
equations}, (Springer series in Nonlinear dynamics, 1994).

\item{5} Golubev V. V.: {\it Lectures on integration of the equations of
motion of a rigid body about a fixed point}, (Moskow: Gostenhizdat, 1953 
 [in Russian]; English translation: Philadelphia: Coronet Books, 1953).

\item{6} Leimanis E.: {\it The general problem of the motion of coupled
rigid bodies about a fixed point}, (Berlin, Heidelberg,
New York: Springer-Verlag, 1965).

\item{7} Manakov S. V.: Remarks on the integrals of the Euler equations 
of the n-dimensional heavy top. {\it Funkc. Anal. Appl.} {\bf 10} (1976 
[in Russian]), 93-94.

\item{8} Nekrasov P. A.: it Analytic investigation of a certain case of
motion of a heavy rigid body about a fixed point. {\it Mat. Sbornik} {\bf 18}
 (1895), 161-274.

\item{9} Ratiu T.: Euler-Poisson equation on Lie algebras and the 
N-dimensional heavy rigid body. {\it American Journal of Math} {\bf 104} 
(1982), 409-448.

\item {10} Ratiu T., van Moerbeke P.: The Lagrange rigid body motion.
{\it Ann. Ins. Fourier, Grenoble} {\bf 32} (1982), 211-234.

\item{11} Reyman A. G., Semenov-Tyan-Shanskiy: Lax representation with 
spectral parameter for Kovalevskaya top and its generalizations.
{\it Funkc. Anal. Appl.} {\bf 22} (1988 [in Russian]), 87-88.

\item{12} Sretenskiy L. N.: On certain cases of motion of a heavy
rigid body with gyroscope. {\it Vestn. Mosk. Univ.} No. 3 (1963 [in
Russian]), 60-71. 

\item{13} Trofimov V. V., Fomenko: {\it Algebra and geometry of integrable 
Hamiltonian differential equations }, (Moscow: Faktorial, 1995 [in Russian]).

\item{14} Zhukovski: Geometrische interpretation des Hess'schen
falles der bewegung eines schweren starren korpers um einen festen
Punkt. {\it Jber. Deutschen Math. Verein.} {\bf 3} (1894), 62-70.

\enddocument